\documentclass[myepj-spec,pdftex]{mySvjour}
\usepackage{graphicx}
\usepackage[pdfpagemode=UseNone]{hyperref}
\usepackage{color}
\usepackage{xspace}
\usepackage[square,sort&compress,numbers]{natbib}   
\usepackage{verbatim}
\usepackage{amsmath,amssymb,amsfonts} 
\usepackage{tabularx}
\usepackage{multicol}
\usepackage{footnote}
\usepackage{listings}
\usepackage[gen]{eurosym}
\usepackage[switch, modulo]{lineno}
\usepackage{lmodern,bm}                
\usepackage[T1]{sansmath} 
\SetMathAlphabet{\mathsfbf}{sans}{\sansmathencoding}{\sfdefault}{bx}{sl}
\usepackage{etoolbox}
\AtBeginEnvironment{sansmath}{}{}{}

\usepackage{lipsum}

\newcommand{\June}{\textsc{June}\xspace}
\newcommand{\JuneGermany}{\textsc{June-Germany}\xspace}

\definecolor{darkblue1}{rgb}{0,0,.2}
\definecolor{darkblue}{rgb}{0,0,.2}
\definecolor{darkred}{rgb}{0.5,0,0}
\pagecolor{white} 
\color{black}     
\hypersetup{breaklinks=true, 
	colorlinks=true, 
	linkcolor=darkblue1, 
	menucolor=darkblue1, 
	urlcolor=darkblue1,
	citecolor=darkblue1,
	pdftitle={},
	pdfauthor={},
	pdfsubject={},
	pdfkeywords={},
	pdfproducer={}
}
\usepackage{siunitx}

\bibstyle{plain}

\begin{document}
	
	\twocolumn[{%
		\begin{@twocolumnfalse}
			
			\begin{flushright}
				\normalsize
			\end{flushright}
			
			\vspace{-2cm}
			
			\title{\Large\boldmath On the Impact of School Closures on COVID-19 Transmission in Germany using an agent-based Simulation}
			%

\author{Lucas Heger\inst{1}, Kerem Akdogan\inst{1} \and Matthias Schott \inst{1,2}}
\institute{%
    \inst{1} Institute of Physics, Johannes Gutenberg University, Mainz, Germany\\
    \inst{2} Corresponding author: schottm@uni-mainz.de
}

			
			\abstract{The effect of school closures on the spread of COVID-19 has been discussed among experts and the general public since those measures have been taken only a few months after the start of the pandemic in 2020. Within this study, the \JuneGermany framework, is used to quantify the impact of school closures in the German state Rhineland Palatinate using an agent-based simulation approach. It was found that the simulations predicts a reduction of the number of infections, hospitalizations as well as death by a factor of 2.5 compared to scenarios, where no school closures are enforced, during the second wave between October 2020 and February 2021.} 
	\maketitle
	\end{@twocolumnfalse}
}]


\section{Introduction}

During the COVID-19 pandemic, countries worldwide implemented a range of measures to mitigate the spread of the virus. These included widespread lockdowns and movement restrictions, social distancing guidelines, quarantine and isolation protocols, mask mandates, and intensive testing and contact tracing efforts. Travel restrictions were imposed, and remote work and online education were encouraged. One of the most controversial non pharmaceutical interventions have been school closures. Faced with the highly contagious nature of the novel coronavirus, many governments around the world temporarily closed educational institutions, ranging from primary schools to universities. These closures aimed to reduce the risk of transmission among students, teachers, and staff, as crowded school settings presented significant challenges for maintaining social distancing. Several aspects on the effect of school closures on the pandemic have been already studied, e.g. \cite{School1, School2, School3, School4}, not even mentioning a wide range of studies on the resulting sociological implications. Most studies find a significant reduction in infections, in particular in the international context, e.g. \cite{RePEc:osf:socarx:v2ef8}. 

Within this study, we conduct the first systematic evaluation on the effect of school closures in the state of Rhineland Palatinate (RLP), Germany, based on a detailed simulation. The mathematical simulation and forecasting of the evolution of epidemics or pandemics is typically based either on compartmental models or on agent based models. The first divide the population into compartments (e.g., susceptible, exposed, infected, recovered) and use differential equations to describe how individuals move between these states (e.g., SIR model \cite{articleWangSir, articleSir2}), while the latter simulate individual agents with specific characteristics and behaviors, allowing for a more detailed representation of interactions and spatial dynamics \cite{Bullock:2021, pillai2023agentbased, muller2020realistic, articleABM1}. Within this work, we are interested in the relative change of infections, hospitalizations and death with and without school closures. An agent based simulation seems in particular suited to answer this question, as the infections can be simulated on individual level. 

The agent-based framework, which is used for our study is briefly described in Section \ref{sec:framework}, followed by a summary of the predictive power of the framework within RLP during the second Covid-19 wave between October 2020 and February 2021 (Section \ref{sec:simRLP}). The effect of school closures is discussed and interpreted in Section \ref{sec:results}, following by a brief conclusion.

\begin{figure*}[thb]
\centering
    \includegraphics[width=0.49\textwidth]{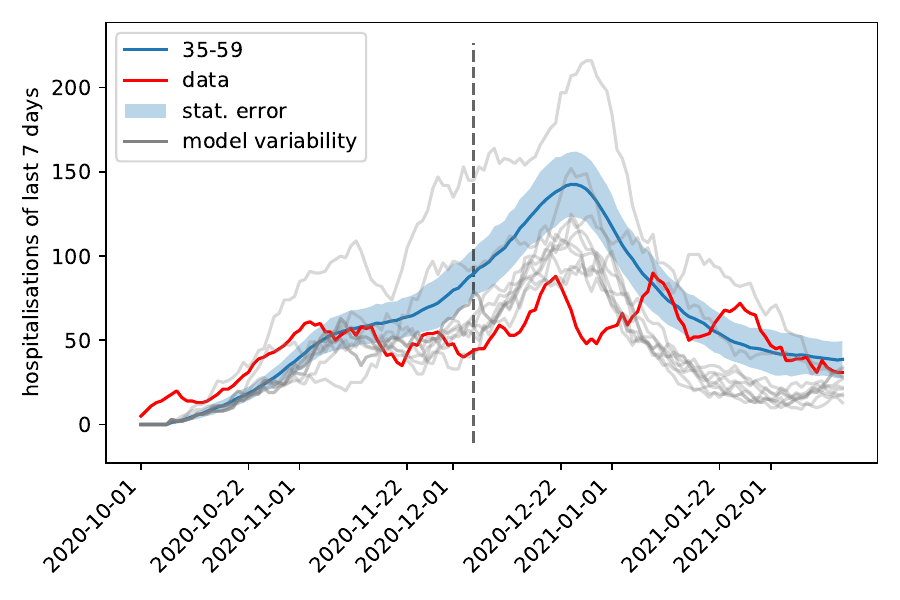}
    \includegraphics[width=0.49\textwidth]{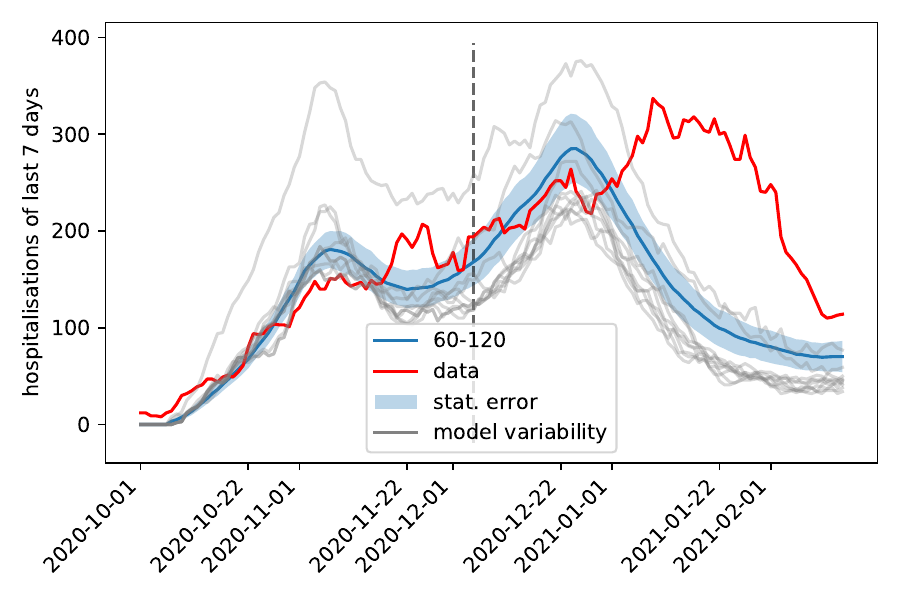}
    \caption{Hospitalization rate of the last 7 days  (red line) for the age group 35-59 (left) and $\geq$60 (right), together with the \JuneGermany simulation with the best fitting parameters (blue line), its statistical uncertainty (shaded blue) and the simulated curves of alternative sets of model parameters tested during the fitting procedure (gray). The vertical line indicates the date until when the data was fitted.}
\label{fig:res1}
\end{figure*}

\section{The \JuneGermany Framework\label{sec:framework}}

The agent-based model known as the \June framework \cite{Bullock:2021} was crafted to simulate epidemics within a population, specifically during the initial and subsequent waves of the Covid-19 pandemic. This model uniquely integrates detailed geographic and sociological data for England, derived from the 2011 Census \cite{Zensus2011}. Significantly, \June allows for a model that accurately forecasted the geographical and sociological dynamics of Covid-19 transmission. A detailed discussion of the \June framework is presented in \cite{Bullock:2021}. 

\JuneGermany \cite{junegermany} is based on the \June framework, however, adapted to Germany. It is also implemented in \textsc{Python} and structured into four interconnected layers: The population layer details individual agents, their static social environments (e.g., households, workplaces), and demographics across hierarchical geographic layers. The interaction layer captures daily routines (commuting, leisure) while the disease layer models disease transmission and effects. Government policies for pandemic mitigation are incorporated in a policy layer. 

The agents within \JuneGermany follow their daily routines in discrete time-steps, and are associated with specific households, schools, and workplaces. The interactions are defined via social contact networks. The disease transmission during public transportation is considered as well as age-dependent social interaction matrices, that model contact frequency/intensity in various settings. The framework utilizes probabilistic infection modeling, considering factors like transmissive probability, susceptibility, and exposure time. Infected agents experience health impacts, from asymptomatic cases over ICU admission to a lethal course . Government policies, at localized levels, consider geographical regions and social interactions, allowing modeling of essential workers' activities and general population compliance. The compliance of the agents to those restriction depend on social and demographic parameters.

The geographical model in this simulation is based on German administrative areas, featuring three layers: states, districts, and municipalities. The 2011 Census \cite{Zensus2011} serves as the geographical basis, offering detailed demographic data, including household composition. The population in each municipality is generated based on age and sex distribution, resulting in varying population densities. The simulation considers age as the most significant risk factor for severe Covid-19 cases, minimizing the impact of co-morbidity variations.

The household compositions are also extracted from the 2011 Census data-set \cite{Zensus2011} and are described by a limited set of categories depending on the number of adults and the number of children living in each household. 

Most importantly, 14,502 primary schools with an average number of 204 students per school are modelled, using public available information at state-level. The secondary schools, of which 13,068 are included in the \JuneGermany framework, have a significantly larger average of students per school, i.e. 506. An average teacher to student ratio of 0.12 is used and a class sizes between 20 and 30 students are assumed. All agents with the corresponding age groups are distributed to the simulated school according to their home address. 

Jobs are classified into sectors using the International Standard Industrial Classification. To model companies in each district per sector, the average number of employees for a company in a specific sector is multiplied by the number of people working in that sector in the district. During population generation within \June, individuals are assigned a workplace in either the super area of residence or a neighboring super area based on mobility data.

The modeling of social activities and interactions is similar in \June and \JuneGermany: agents' weekday routines encompass four distinct activities: work/school, shopping, leisure, and staying at home. Beyond working hours, social activities extend from visits to cinemas and theaters to gatherings with friends in pubs and restaurants, clearly depending on the current state regulations as well as the compliance of those. Commuting is implemented by a directed network graph and considers both, short-distance and long-distance travels. 

\section{Simulation of the COVID-19 Pandemic in the State of Rhineland Palatinate\label{sec:simRLP}}

We use \JuneGermany to model the second wave of the Covid-10 Pandemic in the German state of Rhineland-Palatinate between October 2020 and February 2021. The number of cumulative deaths in the age groups 0-4, 5-14, 15-34, 35-59 and $\geq$60 from the 1st of October 2020 to the 14th December 2021 were used to determine the best model parameters. Most importantly are the cumulative death in the age group above 35, as only very few death have been reported for younger population. 

Due to the complexity of the JUNE simulator, particularly the large dimension of both input parameter and output space, emulation and history matching \cite{hme1, hme2} was used to try to find acceptable matches to data. For this, we used the in-development \textsc{R} package \textsc{hmer} \cite{hmer}, which streamlines the process of constructing emulators and generating representative parameter sets for later iterations of emulation and history matching, and has been used for parameter estimation in other epidemiological scenarios. The emulation and history matching framework offers a number of advantages over traditional methods of parameter estimation, most importantly that relatively few evaluations from the (computationally expensive) simulator to train an emulator are required. Once the optimisation of model parameters has been performed using data up to the 14th of December 2021, we predict the full second wave until the 22nd of February 2022.

The comparison of the hospitalization rate of the last 7 days is shown in Figure \ref{fig:res1}, where a good agreement can be observed, although the simulation predicts a faster decline of the wave than observed in the data. The total number of patients predicted to be hospitalized during the entire second wave is 5181, compared with the official number of 5638. A significantly less precise prediction is observed for the incidence rate. While the number of infections is correctly described for the age group $\geq$60, for the age groups 14-34 and 35- 59 there is a discrepancy by a factor of about three, i.e. a number of unreported cases. This ratio increases significantly for the 5-14 age group. While a general trend towards unreported cases is expected, the extend of the observed underreporting seems surprising, however also observed in other simulation studies, e.g. \cite{pesaran2022matching}. A detailed discussion can be found in \cite{junegermany}.

\begin{figure*}[t!]
\centering
    \includegraphics[width=1.0\linewidth]{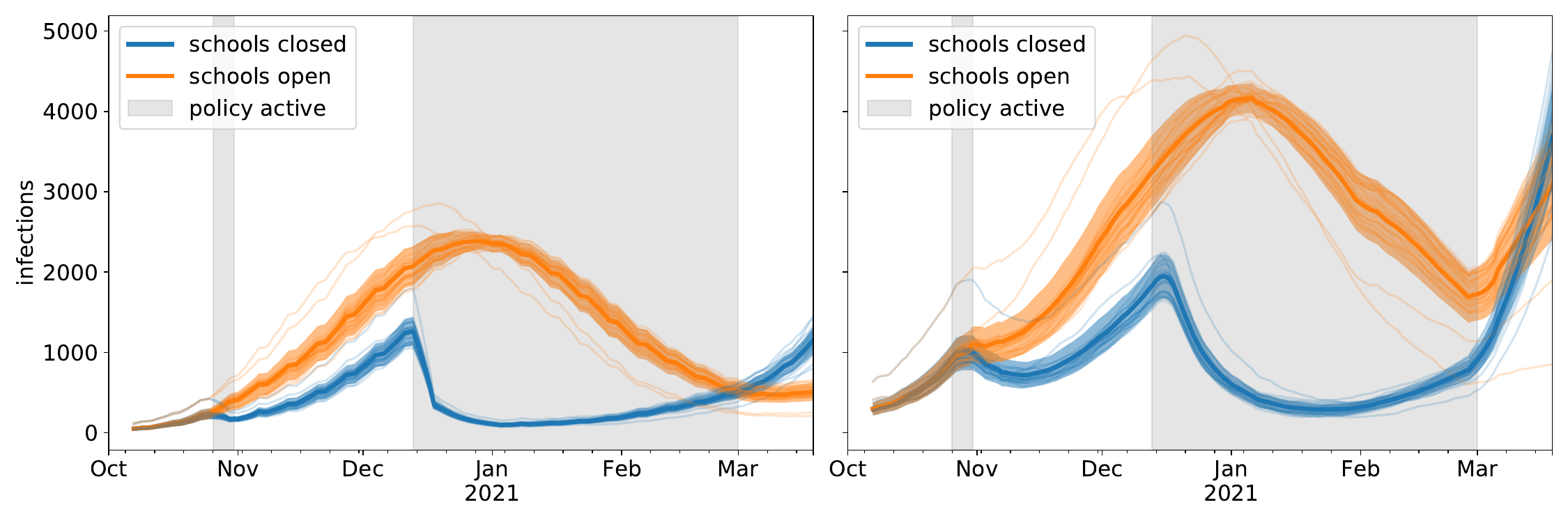}
    \includegraphics[width=1.0\linewidth]{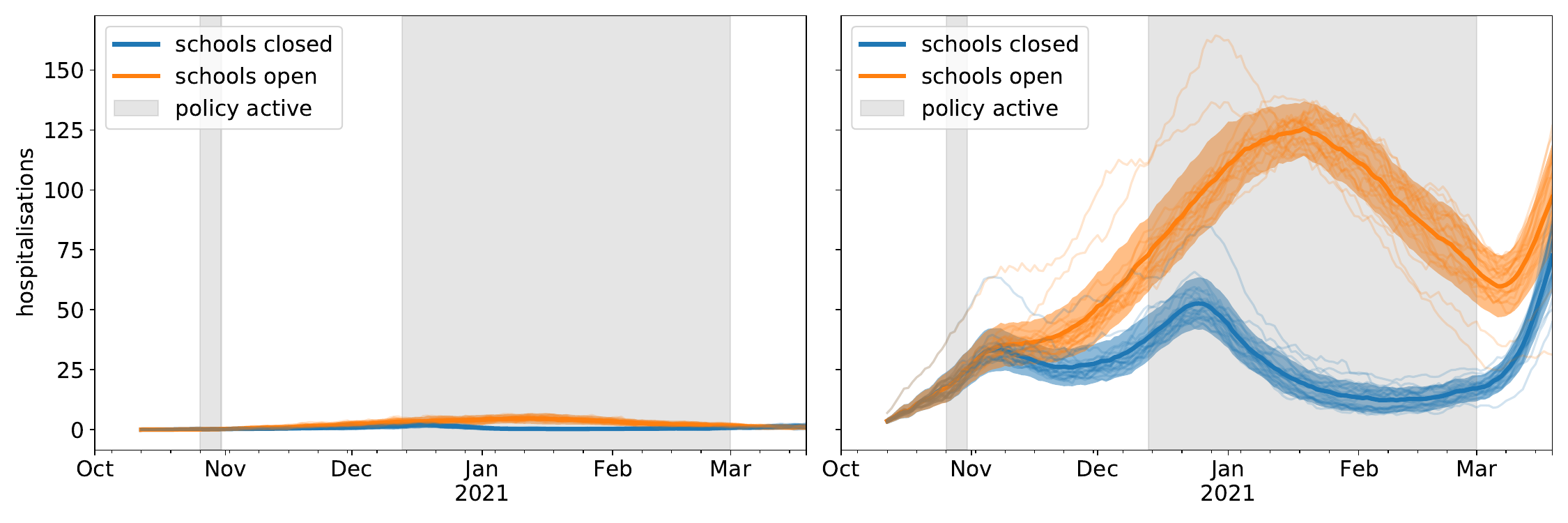}
    \caption{The Figure compares plausible simulations with open schools to simulations where schools are closed. The left plots show the effect among students and the right plot among all other groups in the population. The upper row shows the infection, the lower row shows the hospitalisation rate.}
    \label{fig:open_closed_comparison}
\end{figure*}

\section{\label{sec:results} Study on the effect of School Closures}

The study of the impact of school closures on COVID-19 transmission within RLP is based on the model parameters (and uncertainties) previously discussed. Two sets of simulations have been performed: one with open schools and one with closed schools, altering the policy settings within \JuneGermany, following the official restrictions given during the considered time by the state of RLP. In the second scenario no restrictions on the schools visits are enforced in the simulation, i.e. all relevant agents continue their daily routines regarding school visit. All other restrictions and recommendations are however still applied. 

The analysis of daily infections, hospitalisation rate and deaths serve as a metric to measure the effectiveness of school closures in reducing virus transmission. To establish the reliability of the observed differences between both scenarios, a large part of the parameter space of the model are evaluation, corresponding to parameters that matched the data within a $68\%$ agreement range \cite{junegermany}.

The results of the simulations are shown in Fig.~(\ref{fig:open_closed_comparison}). We examined the number of infections as well as the number of deaths among students (left) and the rest of the population (right), differentiating between the two scenarios. Multiple runs in the parameter space are shown, with the mean being highlighted. We smoothed the infections by choosing a seven-day mean. 

For students, we observed a notable contrast across all metrics, between the two scenarios. In the case of schools being closed from December on wards (depicted in blue on the left side of the figure), the number of infections decreased significantly, reaching a minimum of 100. Conversely, in the scenario with schools remaining open, the number of infections surged to approximately 2400. This stark difference highlights the substantial effect of school closures in mitigating the spread of COVID-19 among pupils.

The impact of school closures extended beyond the student population to the rest of the population, indicating that infections among students, are a significant factor in the dynamic of the pandemic. On the right side of the plot, we observed similar trends in infection rates. When schools remained open, the total number of infections increased considerably. An increase of infections after March for students as well as the general public and can be explained by the fact that also several other governmental policies run out at this point. 

Several things need to be noted: We are comparing two scenarios where only one parameter in the simulation has been changed. Any mis-modeling of the overall infection rate due to the insufficient description of other model parameters will be canceled to first order when taking the ratio of both scenarios. A typical example would be the rate of underreporting, as this affects both scenarios with a constant factor. In particular, when choosing model parameters with school closing policies, the effect on the relative difference when comparing scenarios with and without school closing will typically remain largely unaffected. Other parameters might induce time-dependent differences, but even for those, the ratio will mitigate the effects of mis-modeling.

When comparing all 30 scenarios with and without school closings policies based on model parameter configurations that match the observed hospitalisation, we find an increase by a factor of $2.5\pm0.05$ on the incidence rate. The smallest observed increase was a factor of 2.4, the largest observed increase a factor of 2.7. The increase on the hospitalisation rate and death rate is consistent, yielding factors of $2.6\pm0.2$ and $2.6\pm0.1$ respectively. 


\section{Conclusion}

Our studies indicate a reduction of infections, hospitalization rate and lethal courses during the Covid-19 by factor of 2.5 when primary, secondary schools as well as universities are closed compared to a scenario when no closure is enforced. In particular, we could show  that school closures not only protect pupils but also have a considerable influence on reducing infections within the broader population. These findings underscore the importance of school closures as an effective measure in controlling the transmission of COVID-19, indicating that such interventions can have a significant impact on public health outcomes. Given that Rhineland Palatinate is a medium sized state within Germany with 3.6 million inhabitants, including larger cities and more rural areas, we would argue that the general conclusion should stay largely valid for the whole of Germany. Clearly, we cannot make any statement on the long term social implications of school closures and those effects need to be taken in consideration, when deciding about school closing in potential future epidemics.

\section*{Acknowledgement}

This work has been supported by the Johannes Gutenberg Startup Research Fund. Part of the simulations were conducted using the supercomputer Mogon II at Johannes Gutenberg University Mainz (hpc.uni-mainz.de). The authors gratefully acknowledge the computing time granted on the supercomputer.



\bibliographystyle{myelsarticle-num}
\bibliography{./Bibliography}

\end{document}